\documentclass[12pt]{article}

\usepackage{times}
\usepackage[T1]{fontenc}
\normalfont
\usepackage{mathptmx}
\usepackage{graphicx}

\newcommand{\lsim}{\mbox{\raisebox{-1.ex}{$\stackrel
     {\textstyle<}{\textstyle\sim}$}}}
\newcommand{\gsim}{\mbox{\raisebox{-1.ex}{$\stackrel
     {\textstyle>}{\textstyle\sim}$}}}

\usepackage{float}
\usepackage{amssymb}
\usepackage{amsmath}

\begin{document}

\nopagebreak

\baselineskip=16pt

\begin{titlepage}

\begin{flushright}
SIAS-CMTP-06-03\\[-1mm]
\end{flushright}
\begin{center}
\vspace*{1.2cm}

{\Large\bf 
A Holographic Dual of CFT with Flavor on de Sitter Space
}\vspace{1cm}

Takayuki Hirayama\vspace{5mm} 

{\it Shanghai Institute for Advanced Studies,
University of Science and Technology of China,
\vspace*{-0.5ex}\\
99 Xiupu Road, Pudong, Shanghai, 201315, China}
\vspace{10mm}

\end{center}

\begin{abstract}\noindent
 We introduce a D7-brane probe in $AdS_5\times S^5$ background in a way
 that the 4d part of the induced metric on D7-brane becomes 4d de-Sitter
 space ($dS_4$) inside $AdS_5$ instead of 4d Minkowski space. Although
 supersymmetry is completely broken, we obtain a static configuration and
 show the absence of dynamical tachyonic modes. Following holographic
 renormalization we  renormalize the Dirac-Born-Infeld action of
 D7-brane and we completely fix the counter terms including finite
 contributions from the consistency under various limits. Through the
 AdS/CFT correspondence we study the chiral condensate and meson
 spectrum of CFT dual theory on $dS_4$ where the energy scale is
 identified with the direction normal to $dS_4$ space in $AdS_5$. We
 identify and properly reproduce the finite temperature effects on
 $dS_4$. Our results support the holographic interpretation of the
 Randall-Sundrum model with non fine-tuned $dS_4$ brane(s) and the
 holography between $AdS_p$ (or $dS_p$) bulk gravity and CFT on
 $dS_{p-1}$ called the (A)dS/dS correspondence.
\end{abstract}

\footnotetext{hirayama@ustc.edu.cn}

\end{titlepage}
\setcounter{footnote}{0}

\pagebreak

\section{Introduction}

The holographic principle conjectured by 't Hooft~\cite{'tHooft:1993gx}
and Susskind~\cite{Susskind:1994vu} is one of the most important keys to
understand and construct quantum theory of gravity. This proposal
deeply concerns the fundamental principles of spacetime and quantum
theory. The explicit example of holography was realized in string theory
with D-branes by Maldacena~\cite{Maldacena:1997re} and is called the
AdS/CFT correspondence. The best understood example of the
correspondence is supergravity (typeIIB string theory) on
$AdS_5\times S^5$ is dual to 4d N=4 super Yang-Mills theory (SYM). A
field on $AdS$ is identified with an operator in CFT and the
generating functional of correlators in CFT is calculated from
$AdS$ gravity~\cite{Gubser:1998bc}. 

Many deformations of AdS$_5$/CFT$_4$ have been discussed by
compactifying one or more spacetime directions, replacing 
%%%an extremal D$p$-brane 
$AdS$ by asymptotically $AdS$ geometries and so on, and
the parameters describing modifications correspond to the temperature or
coupling constants which may break some of supersymmetry in CFT
side (see~\cite{Witten:1998zw} and also \cite{Aharony:1999ti} for a
review).

The holographic principle has been also conjectured for other spacetime
backgrounds 
which have not been yet realized or well understood in string theory.
Strominger 
%%%proposed 
discusses the dS/CFT correspondence~\cite{Strominger:2001pn},
and a Minkowski/CFT correspondence is also
%%%discussed
proposed~\cite{deBoer:2003vf}. The 
%%%CFT 
holographic interpretation of the
Randall-Sundrum model~\cite{Randall:1999ee} was developed
in~\cite{Verlinde:1999fy}.
Hawking-Maldacena-Strominger~\cite{Hawking:2000da} speculate the AdS/CFT
interpretation of the Randall-Sundrum model with non fine-tuned branes,
i.e. $AdS_5$ space bounded by $dS_4$ branes, by deriving the
same value of the entropy from $AdS$ gravity and from CFT on a $dS$
space.
Alishahiha et al.~\cite{Alishahiha:2004md, Alishahiha:2005dj} discuss
a broader holography between $dS_p$ or $AdS_p$ bulk gravity and CFT on
$dS_{p-1}$ which is realized as a slice of bulk geometry, the (A)dS/dS
correspondence (see also~\cite{Buchel:2002wf} for other attempts to
realize curved spaces). 

However since the explicit string realization of these spacetime
backgrounds has not been well understood, many of these arguments are
limited to be qualitative. Moreover in the last two examples, the
holography
is in some sense unexpected since the same $AdS$ geometry has different
CFT dual descriptions depending on how $AdS$ geometry is sliced,
a Minkowski slice or a de-Sitter slice.

A classical configuration of a bulk field in $AdS_5$ contains the
information of correlation function of the corresponding operator in
CFT. Let us discuss a
static configuration along 4d Minkowski space. The configuration is
thus a function of normal direction to 4d Minkowski space. There
are two independent integration constants in the asymptotic solutions
near the $AdS$ boundary. They correspond to the coupling constant and
the vacuum expectation value of the corresponding CFT operator.
Those are, of course, some numerical numbers. Now let us apply the same
logic to a static configuration along $dS_4$ space in the (A)dS/dS
correspondence. (In this paper we only consider $AdS_p$ bulk and do not
consider $dS_p$ bulk geometry.) The static configuration is a function
of normal direction to $dS_4$ which is NOT a normal direction to 4d
Minkowski space, but a combination of that and time direction of 4d
Minkowski space. Using the original AdS/CFT correspondence, we
obtain a time dependent coupling constant and a time dependent vacuum
expectation value of the same CFT operator (Section~5). Although we are
studying the same theory, we are looking at a 
sector where 4d Lorentz symmetry is complicatedly broken. So what is
this sector? The guide is the symmetry. Since the static configuration
is independent of $dS_4$ coordinates, the coupling constant and vacuum
expectation value are interpreted as those on $dS_4$. We will explain
this idea in Section~5 and speculate a conformal transformation relates
the (A)dS/dS correspondence ($AdS_p$ bulk) with the AdS/CFT
correspondence.

Without string realization, the above argument just tells a possibility
that a different slice of $AdS$ gives a
different result on the dual theory. In this paper we realize the setup of
the (A)dS/dS correspondence in string theory in a probe limit.
The gravity dual of adding small number of dynamical quarks into N=4 SYM
has been proposed by Karch and Katz by introducing a D7-brane as a
probe in the D3-brane background 
$AdS_5\times S^5$~\cite{Karch:2002sh}. We
introduce a probe D7-brane in a different way such that the probe
D7-brane fills $dS_4$ inside $AdS_5$ and other directions except the
$S^2$ inside the $S^5$ (Section~2). The locus in the $S^2$ is described
by a scalar field localized on D7-brane and is a function of the normal
direction to $dS_4$ surface in $AdS_5$. In the CFT side this corresponds
to introducing fundamental quarks with supersymmetry breaking
mass terms (see Section~5). We find a stable
supergravity configuration and this realizes the setup of the (A)dS/dS
correspondence, since the scalar field is a bulk field in $AdS_5$ and is
a function of the direction normal to $dS_4$. Then applying the AdS/CFT
dictionary~\cite{Gubser:1998bc} and holographic
renormalization~\cite{Henningson:1998gx, Balasubramanian:1999re}, 
we show the corresponding CFT operator is the quark anti-quark
composite operator as it is in~\cite{Karch:2002sh} and study
the flavor physics which are the
chiral condensate (Section~2), meson spectrum (Section~3) and quark
anti-quark potential (Section~4). We then discuss whether the flavor
physics we obtain is consistently understood as that in CFT with flavor
on $dS_4$.

The $dS_4$ space has a temperature proportional to the inverse of
curvature radius. We expect the finite temperature effects on flavor
physics do not severely depend on the detail of large N gauge theory,
and are qualitatively same as those in a large N gauge theory with
flavor on Minkowski space at finite temperature.
%%%%%%%%%%%%%%%%%%
%%%  modified  %%%
%%%%%%%%%%%%%%%%%%
In fact many works on flavor physics using the gravity dual of
nonsupersymmetric large N gauge
theories~\cite{Kruczenski:2003uq, Evans:2004ia} have shown that the
qualitative behaviour of flavor physics does not depend on the detail of
large N gauge theory. It has been also known that the finite
temperature effects on flavor physics are insensitive to the detail of
large N gauge theory~\cite{Roberts:2000aa, Babington:2003vm,
Kruczenski:2003uq, Ghoroku:2005tf}. 
%For example, the following
%qualitative behaviours
%are universal: the phase is divided into the confinement and
%deconfinement phases depending on the ratio between the quark mass and
%temperature, and the mass gap between mesons becomes zero in
%the deconfiment phase.

Therefore we will thus identify the finite temperature effects on flavor
physics in our calculations and see whether they are properly
reproduced. We will show that they are indeed properly reproduced. Thus
we believe our results give an support on the holographic understanding
of the Randall-Sundrum model with non fine-tuned $dS_4$ brane(s) and the
(A)dS/dS correspondence.

\section{Probe D7-brane and the chiral condensate}

The effective theory on the supersymmetric D3/D7 system is known to be
N=2 SYM with
charged hypermultiplets which are incorporated from the D3-D7 open
strings. Karch and Katz proposed the corresponding supergravity solution
by treating D7-brane as a probe\footnote{
The backreaction of a probe is discussed
in~\cite{Erdmenger:2004dk}.} 
in $AdS_5\times S^5$
geometry~\cite{Karch:2002sh}. In this paper we instead embed a D7-brane
probe in a way that supersymmetry is completely broken. The $AdS_5$
space realizes a $dS_4$ space as a
hypersurface in addition to 4d Minkowski space and we embed a probe
D7-brane along this $dS_4$ space (thus supersymmetry is completely
broken) and the normal direction to the $dS_4$ hypersurface in $AdS_5$
as well as
the $S^3$ inside $S^5$. The position of D7-brane in the remaining $S^2$
directions inside $S^5$ is a function of the normal direction to
$dS_4$. In the CFT side, introducing D7-brane corresponds to adding
fundamental quarks and all the results we will compute suggest that we
add fundamental quarks on $dS_4$ space.

The D3-brane background is described by $AdS_5\times S^5$
(Appendix~\ref{A})
\begin{eqnarray}
 ds^2&=& \frac{R^2}{l_5^2}dx_{M4}^2 +\frac{l_5^2}{R^2}dR^2 
  +l_5^2d\Omega_5^2
  \label{met1}\\
 &=& \left(\frac{u^2}{l_5^2}-\frac{l_5^2}{l_4^2}\right)dx_{dS4}^2 
  +\left(\frac{u^2}{l_5^2}-\frac{l_5^2}{l_4^2}\right)^{-1}du^2 
  +l_5^2d\Omega_5^2
  \label{met2}
\end{eqnarray}
where 4d Minkowski space (4d de-Sitter space) is apparent in the first
(second) expression. $d\Omega_n^2$ is the metric of $S^n$.
We explicitly write the $AdS_5$ and $dS_4$ curvature radius which
are $l_5$ and $l_4$ respectively. $R$ is the distance from D3-brane,
$R^2=(X^4)^2+\cdots (X^9)^2$ where $X^0 \cdots X^9$ are the rectangular
coordinates. As approaching to the $AdS_5$ boundary,
$u\rightarrow\infty$, these two expressions are same at the leading
order and this is understood that the $dS_4$ curvature effects are
ignored in high energy limit in the gauge theory side.

Now we embed a $dS_4$ space filling D7-brane as a probe in this
geometry. For the computational convenience, we change the
coordinate $u$ into $r$ by
\begin{eqnarray}
 \frac{4r}{l_5}=\left(\sqrt{\frac{u}{l_5}+\frac{l_5}{l_4}}+
  \sqrt{\frac{u}{l_5}-\frac{l_5}{l_4}}\right)^2
\end{eqnarray}
and the $AdS_5\times S^5$ metric becomes
\begin{eqnarray}
 ds^2 &=&r^2\left(\frac{1}{l_5}-\frac{l_5^3}{4r^2l_4^2}\right)^2dx_{dS4}^2 
  +\frac{l_5^2}{r^2}(dr^2+r^2d\Omega_5^2) .
  \label{revmet1}
\end{eqnarray}
We further rewrite 
$dr^2+r^2d\Omega_5^2=d\rho^2+\rho^2d\Omega^2_3+dy^2+dz^2$ and then
\begin{eqnarray}
 ds^2 &=&r^2\left(\frac{1}{l_5}-\frac{l_5^3}{4r^2l_4^2}\right)^2dx_{dS4}^2 
  +\frac{l_5^2}{r^2}(
  d\rho^2+\rho^2d\Omega^2_3+dy^2+dz^2),
  \hspace{3ex}
  r^2=\rho^2+y^2+z^2 .
\end{eqnarray}
In this coordinate system, the world volume of $dS_4$ space filling
D7-brane is embedded with $y=y(\rho)$ and $z=0$. Because of the
rotational symmetry, we place the brane at $z=0$ and the position along
$y$ becomes a non trivial function of $\rho$.

Since D7-brane breaks supersymmetry completely, this static
configuration may have tachyonic directions\footnote{
Some of readers may wonder if there is a static embedding. Since the
regularity of $y(\rho)$ gives only one condition for a second order
differential equation for $y(\rho)$, we always have a static solution.
}
and we should include the time dependence. However as long as the energy
scale associated with the motion of D7-brane (the scale of tachyon
condensation) is much smaller than the
scales which we are interested in, such as $l_5^{-1}$, $l_4^{-1}$ and
the distance between D3 and D7 branes which will be understood as the
mass for charged matter fields, we expect that the AdS/CFT
correspondence is still applicable and the instability of D7-brane
embedding should be interpreted as an instability in the field theory
side. (Actually we will show there are no dynamical tachyonic modes.)

The Dirac-Born-Infield action of D7-brane with the tension $T_7$ and the
equation of motion for $y(\rho)$ are
\begin{eqnarray}
 S &=& -T_7\int d^8x \sqrt{-h}
 \propto\int d\rho \left(1-\frac{1}{4r^2l_4^2}\right)^4
  \rho^3\sqrt{1+y'(\rho)^2} ,
  \\
 y''(\rho)&=&-3\frac{(1+y'(\rho)^2)}{\rho}y'(\rho)
  +\frac{2}{l_4^2r^2-\frac{1}{4}}\frac{y(\rho)-\rho y'(\rho)}{r^2}
  (1+y'(\rho)^2)
  \label{equ}
\end{eqnarray}
where $\sqrt{-h}$ is the determinant of metric and we hereafter use
$l_5=1$ unit. The D7-brane configuration is given by a regular solution
of the equation of motion up to $\rho=0$ or up to the horizon, otherwise
the approximation of using supergravity breaks down and
there is no interpretation of gauge theory dual since $r$ direction 
is related with the energy scale.
%%%%%%%%%%%%%%%%%%
%%%  modified  %%%
%%%%%%%%%%%%%%%%%%
When we
separate one D3 brane from others along $r$ direction, the mass $M_W$ of
gauge boson associated with symmetry breaking
is calculated from the tension of string stretched between the probe D3
brane and the horizon~\cite{Alishahiha:2004md},
\begin{eqnarray}
 M_W&\propto&\frac{1}{\alpha'}\int^r_{r_h}dr\sqrt{g_{00}}\sqrt{g_{rr}}
  =\frac{1}{\alpha'}\left( r-r_h+\frac{1}{4rl_4^2}
		     -\frac{1}{4r_hl_4^2}\right)
\end{eqnarray}
where $r$ and $r_h$ are the position of probe D3 brane and horizon in
(\ref{revmet1}) and $1/\alpha'$ is the string tension. Therefore the
proper distance from the horizon along $r$ direction is related with
the energy scale in the dual field theory~\cite{Alishahiha:2004md}.

\begin{figure}%[H]
 \begin{minipage}{0.5\hsize}
 \begin{minipage}{0.95\hsize}
  \hspace{-3ex}
   \includegraphics[width=90mm]{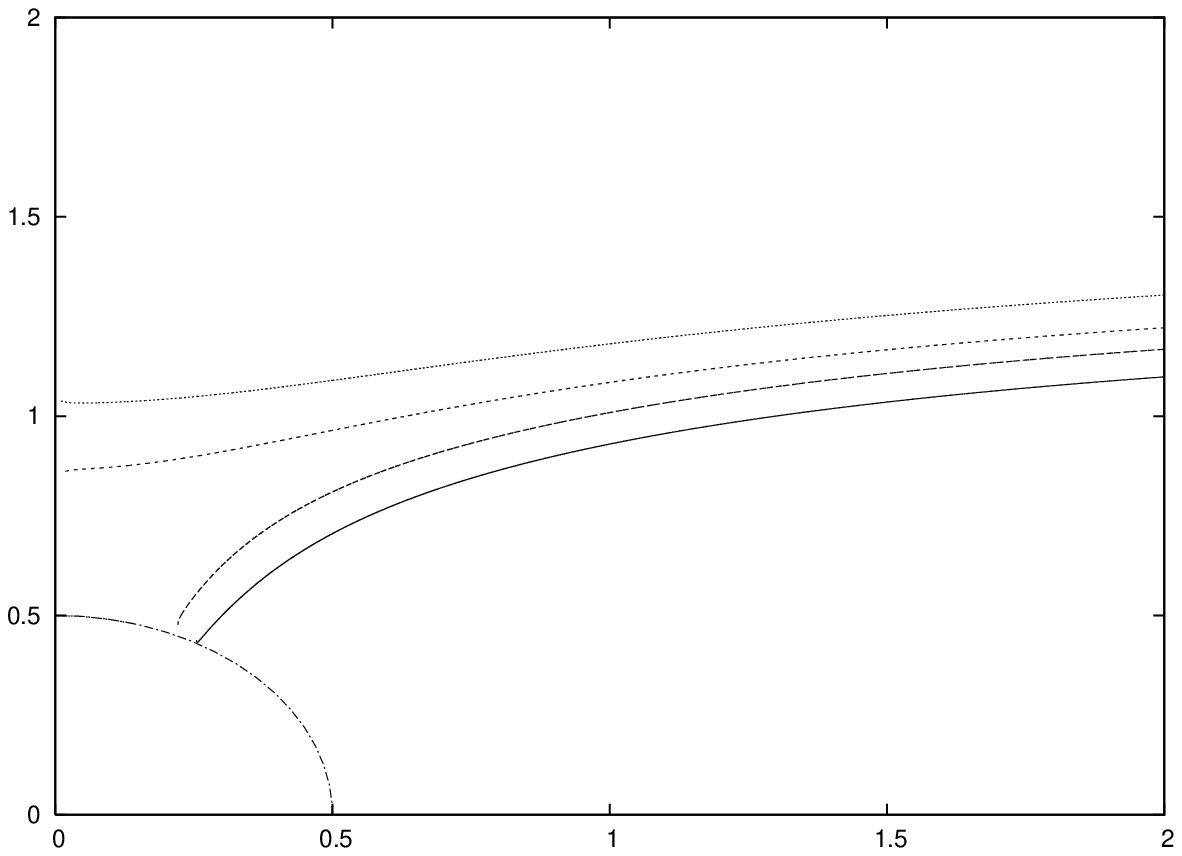}
  \vspace{-3ex}
 \caption{The regular solutions $y(\rho)$ vs $\rho$.
  From the top $m=1.5,1.43,1.37$ and $1.3$. The horizon radius is
 $0.5$. $l_4=1$.}
  \label{fig7}
 \end{minipage}
 \end{minipage}
 \begin{minipage}{0.5\hsize}
 \begin{minipage}{0.95\hsize}
  \vspace{-2.5ex}
  \hspace{-2.5ex}
   \includegraphics[width=90mm]{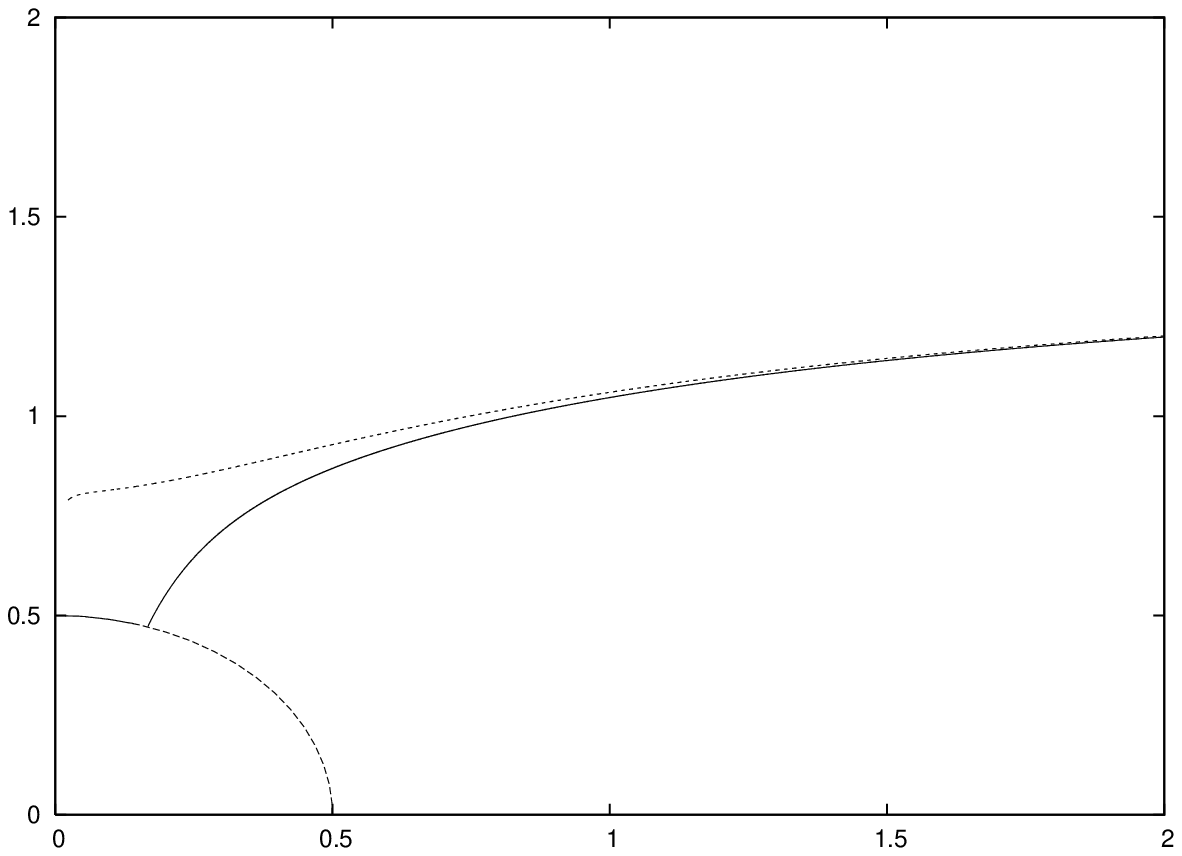}
  \vspace{-3ex}
 \caption{The multiple solutions $y(\rho)$ at $m=1.4$ vs $\rho$.
   $l_4=1$.}
  \label{fig8}
 \end{minipage}
 \end{minipage}
\end{figure}

In the AdS/CFT correspondence the correlation functions of CFT operators
are computed from the on-shell supergravity action which is obtained
from plugging a solution. The on-shell action is usually divergent and
we should regularize and renormalize the action to have a well
defined action. Therefore we first study asymptotic solutions and
divergences in the on-shell action. The asymptotic solutions at
$\rho\rightarrow \infty$ are
\begin{eqnarray}
 y(\rho) &=& m\left(1-\frac{\ln(\rho^2+m^2)}{2l_4^2\rho^2}\right) +
  \frac{v}{\rho^2} +O(\rho^{-4}) .
  \label{asym}
\end{eqnarray}
Though $v$ and $m$ are integration constants in this order, $v$ is
determined in terms of $m$ to give a regular solution,
i.e. $y(\rho)$ and $y'(\rho)$ are everywhere finite.
Plugging the asymptotic solution into the Dirac-Born-Infield action, we
obtain
\begin{eqnarray}
 S &=& -T_7\Omega_3\int d^4x \sqrt{-g_{dS4}}
  \left\{ \frac{\rho_\infty^4}{4}
   -\frac{\rho_\infty^2}{2l_4^2}
   +\left(\frac{3}{8l_4^4}  +\frac{m^2}{l_4^2}\right)\ln\rho_\infty 
   \right\}+S_f
\end{eqnarray}
where $\Omega_3$ is the volume of $S^3$ and $g_{dS4}$ is the $dS_4$
metric. We regularize the action by
introducing the cutoff $\rho_\infty$, which will be taken infinity after
renormalization and $S_f$ is the finite part. We notice that when $m$
is smaller than roughly $1/l_4$ ($1.4/l_4$ from numerical analysis, see
some regular solutions in Fig.\ref{fig7},
% in Appendix~\ref{B}),
a regular solution $y$ 
%%%crosses 
ends at the horizon, and it is easily checked
that there is no divergent contributions to the on-shell action from the
horizon. The
coefficient in front of $\ln\rho_\infty$ is understood as
the conformal anomaly of CFT~\cite{Henningson:1998gx} and is same as
that calculated in~\cite{Alishahiha:2005dj}. Since the $m$ dependent
term in the potential is unbounded below, the mode associated with
changing
$m$ is a 'tachyonic' mode. However the change of potential energy is
(log) divergent and thus this 'tachyonic' mode is not a dynamical mode
(an infinite energy is needed to turn on this mode).
After we introduce counter
terms and renormalize the action, the finite on-shell action would depend
on $m$. However we can impose the Dirichlet type boundary condition for
$y(\rho_\infty)$ as is done in AdS/CFT and fix $m$, i.e. we can
freeze the instability mode. We will later study there is no tachyonic
mode in local excitations. Thus we apply the AdS/CFT correspondence up
to the low energy limit in the field theory side. 

Although in this embedding introducing the cutoff at constant $\rho$
seems natural, this is the key. We do not introduce the
cutoff at constant $R$ in the metric (\ref{met1}). We just simply
follow the AdS/CFT dictionary~\cite{Gubser:1998bc} and holographic
renormalization~\cite{Henningson:1998gx, Balasubramanian:1999re}
with replacing $R$ by $\rho$, and compute correlation
functions in the dual theory. Then we will check whether the resultant
correlation functions are
consistently understood as those in CFT with flavor on $dS_4$.

Since the symmetry of action is useful for determining the possible
counter terms, we notice the invariance of metric under the following
redefinition parametrized by an arbitrary real parameter $\alpha$
\begin{eqnarray}
 l_4\rightarrow \alpha l_4,\hspace{3ex} 
  \rho\rightarrow\alpha^{-1}\rho,\hspace{3ex} 
  y\rightarrow\alpha^{-1}y .
  \label{cf}
\end{eqnarray}
%%%%%%%%%%%%%%%%%%
%%%  modified  %%%
%%%%%%%%%%%%%%%%%%
Therefore although we have two parameters $l_4$ and $m$ in the
asymptotic solutions (\ref{asym}) after taking $l_5=1$ unit, the real
free parameter is only the combination $m l_4\propto m/T$ where $T$ is
the temperature of $dS_4$ space. However we treat both $l_4$ and $m$ as
parameters since it is convenient for discussing the behaviour under the
change of $l_4$ and $m$. 

Following the holographic 
renormalization~\cite{Henningson:1998gx, Balasubramanian:1999re}, we
introduce the counter terms which should respect
unbroken symmetries and are defined at the cutoff
$\rho=\rho_\infty$. Those are
\begin{eqnarray}
 S_c &=& T_7\Omega_3\int d^4x \sqrt{-\gamma} 
  \left. F\left(\frac{y^2}{\rho^2},\rho^2 l_4^2\right)
  \right|_{\rho=\rho_\infty, y=y(\rho_\infty)} ,
  \\
 \sqrt{-\gamma}&=&\sqrt{-g_{dS4}}r^4\left(1-\frac{1}{4r^2l_4^2}\right)^4
  \label{induced}
\end{eqnarray}
where $\gamma$ is the induced metric at a constant $\rho=\rho_\infty$
surface in $AdS_5$ and $y$ appears in the combination $y/\rho$ since
$y/\rho$ is the canonically normalized field in $AdS_5$ and invariant
under (\ref{cf}), and so is $\rho l_4$. The form of $F$ is determined
to cancel the divergences in $S$ and we obtain
\begin{eqnarray}
 S_c &=& T_7\Omega_3\int d^4x \sqrt{-\gamma} 
  \left.\left[
   \frac{1}{4}
   -\frac{y^2}{2\rho^2}
   -\frac{1}{4\rho^2 l_4^2}
   +\left(\frac{3}{8l_4^4\rho^4}+\frac{y^2}{l_4^2\rho^4}\right)
   \left\{(c_1+1)\ln(l_4\rho)+c_1\ln\left(\frac{y}{\rho}\right)\right\}
   \right.\right.
   \nonumber\\
 &&\qquad\left.\left.
   +c_2\frac{y^4}{\rho^4}
   +c_3\frac{1}{\rho^4 l_4^4}
   +c_4\frac{y^2}{\rho^4 l_4^2}
		\right]\right|_{\rho=\rho_\infty, y=y(\rho_\infty)} .
 \label{abc}
\end{eqnarray}
With these coefficients, all the divergences are cancelled out,
i.e. $S+S_c$ is finite. As is discussed in the
literatures~\cite{Henningson:1998gx, Balasubramanian:1999re,
Bianchi:2001de}, we have ambiguities in finite contributions
$c_1 \cdots c_4$ which are numerical constants and do not depend on
$l_4$. A different choice of finite counter terms corresponds
to a different regularization scheme in the field theory side and the
correlation
functions depend on a choice of scheme. Thus we need to find a proper
regularization scheme. We will show that the finite counter terms
are completely fixed from the requirements that the correlation
functions should recover those in supersymmetric theories after taking
the limit $l_4\rightarrow\infty$ and should also recover physically
acceptable ones in the limit where $dS_4$ curvature is negligible.

Since the metric (\ref{met2}) approaches to the metric (\ref{met1}) near
the $AdS$ boundary, we identify
the CFT operator ${\cal O}$ for $y$ field as the same operator in the
supersymmetric case, i.e. the quark anti-quark composite operator
$q\bar{q}$ where quark is a fermionic field in a hypermultiplet.
According to the AdS/CFT dictionary~\cite{Gubser:1998bc}, 
$y(\rho_\infty\rightarrow\infty)=m$ is the mass for quark, i.e.
${\cal L}={\cal L}_{CFT}+m\bar{q}q$, and the one
point function $\langle{\cal O}\rangle$, the chiral condensate, is
calculated from the renormalized action,
\begin{eqnarray}
 \langle {\cal O} \rangle &=&
  \lim_{\rho_\infty\rightarrow\infty}
  \frac{\rho_\infty^4}{\sqrt{-\gamma}}
  \frac{\delta (S+S_c)}{\delta y(\rho_\infty)}
  \\
 &=&  \lim_{\rho_\infty\rightarrow\infty}
  T_7\Omega_3
  \left[ \left\{-\frac{2m}{l_4^2}\ln\rho_\infty+\frac{m}{l_4^2}
	  +2v\right\}
  +\left\{\frac{2m}{l_4^2}\ln\rho_\infty+\frac{m}{l_4^2}(A+B\ln m)
   +Cm^3\right\}
  \right] ,
  \label{finitect}\\
 &&\qquad
  A=2(c_1+1)\ln l_4+c_1+2c_4,\hspace{3ex}
  B=2c_1,\hspace{3ex}
  C=4c_2 ,
\end{eqnarray}
where $\gamma$ is the induced metric at a constant $\rho=\rho_\infty$
as (\ref{induced}). The terms inside the first bracket come from the
action $S$ and those inside the second bracket come from the counter
terms $S_c$. The log divergent terms cancel out and finite counter terms
do contribute to the one point function. In the supersymmetric case,
i.e. $l_4\rightarrow \infty$, $2v+Cm^3$ survive. Since the
regularity of solution is satisfied only for $v=0$ and the total
on-shell value of action should be zero because of
supersymmetry~\cite{Bianchi:2001de}, $C=0$
follows and there is no condensate. When $l_4$ is finite, $v$ in a
regular solution is a complicated function of $m$. We expect the chiral
condensate $\langle {\cal O} \rangle$ goes to zero as quark mass
gets large $m\gg 1/l_4$ because the $dS_4$ curvature effects should be
negligible, and thus other coefficients $A$ and $B$ should be
determined such that  $\langle {\cal O} \rangle$ goes to zero as
$ml_4\rightarrow\infty$.
If we cannot realize $\langle {\cal O} \rangle\rightarrow 0$
in this limit, it indicates the AdS/CFT correspondence has to be modified
or cannot apply for this case. Therefore this gives a nontrivial check of
the AdS/CFT correspondence in our generalized cases.

%%%%%%%%%%%%%%%%%%
%%%  modified  %%%
%%%%%%%%%%%%%%%%%%
The ambiguities of finite counter terms exist in the case of other
geometries as well. For example in the case of $AdS$ Schwartzschild
black hole studied in~\cite{Babington:2003vm}, there is one free
parameter labells the finite counter term which corresponds to 
$C$ in~(\ref{finitect}). This free parameter is fixed such that the
chiral condensate goes to zero at the supersymmetric limit or
a large quark mass $m/T\rightarrow\infty$ ($T$ is the black hole
temperature) limit and $C=0$ is obtained. Thus 
the chiral condensate in~\cite{Babington:2003vm} is just given by $v$ in
our notation.

\begin{figure}
 \begin{minipage}{0.5\hsize}
 \begin{minipage}{0.95\hsize}
  \hspace{-2.5ex}
   \includegraphics[width=90mm]{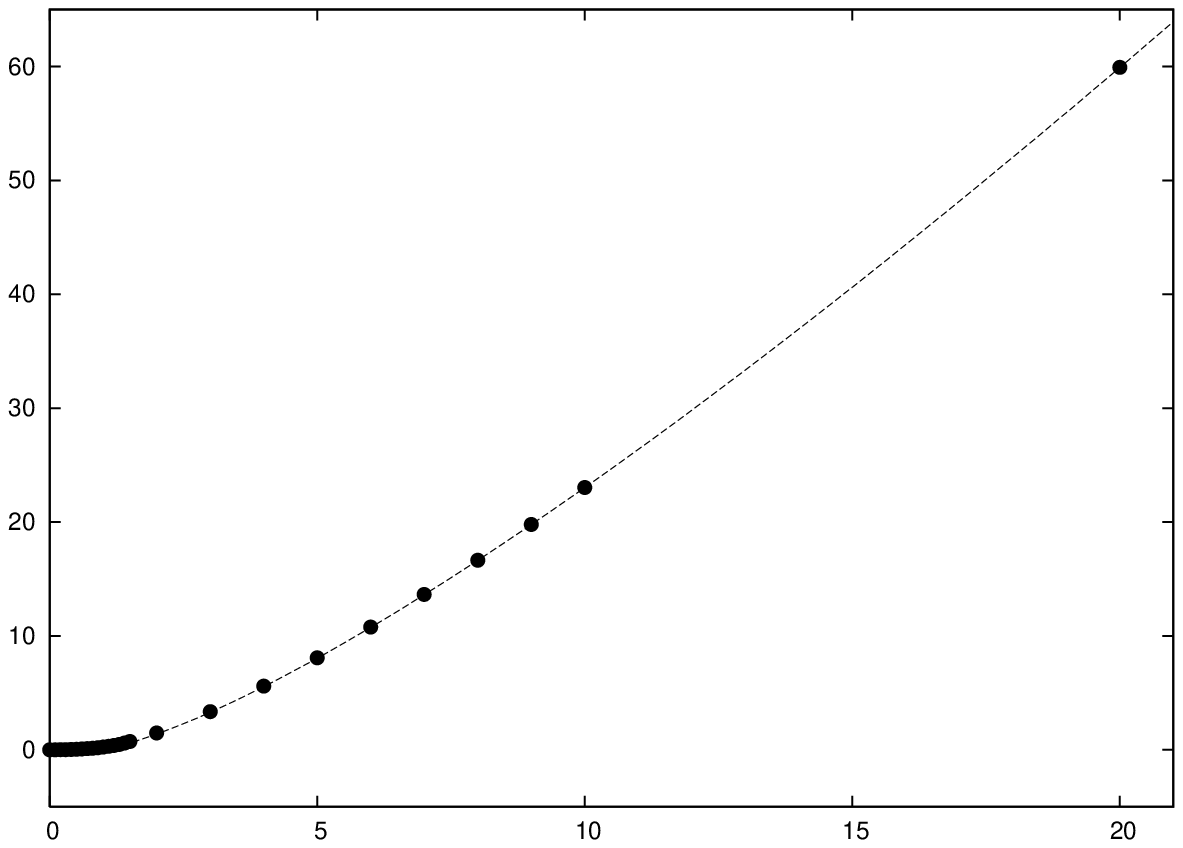}
  \vspace{-4ex}
  \hspace*{2.5ex}
 \caption{$v$ of regular solutions vs $m$.
  The line is the best fit function 
%$-1.7\times 10^{-4}m+1.0m\ln m$
  $1.0\times m\ln m$
  which almost overwraps the dots in this scale. $l_4=1$.}
  \label{fig1}
 \end{minipage}
 \end{minipage}
 \begin{minipage}{0.5\hsize}
 \begin{minipage}{0.95\hsize}
  %\vspace{-1.5ex}
  \hspace{-2.5ex}
   \includegraphics[width=90mm]{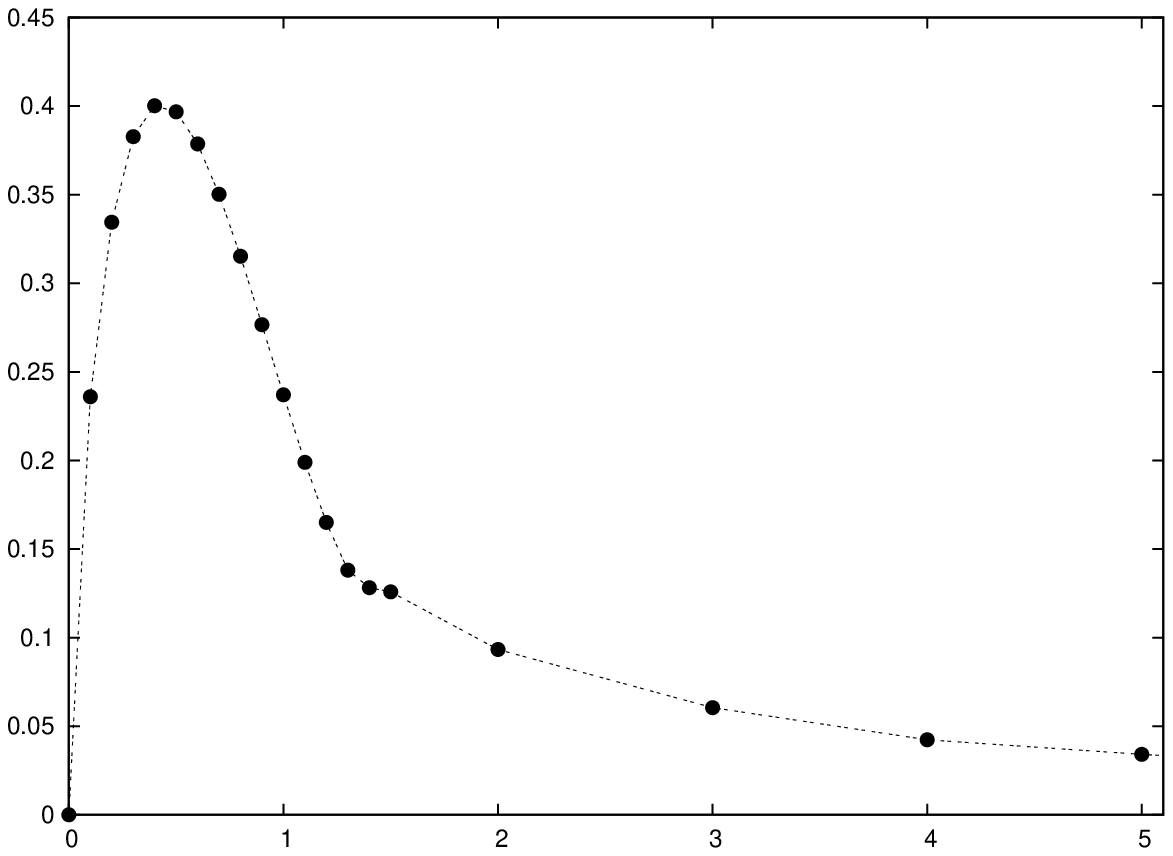}
  \vspace{-4ex}
 \caption{The chiral condensate $\langle {\cal O}\rangle$ vs $m$.
  $l_4=1$. (The small difference between the line and numerical data in
  Fig.\ref{fig1} gives this.)}
  \label{fig2}
 \end{minipage}
 \end{minipage}
\end{figure}

We solve the equation of motion numerically and find regular solutions
of $y$ by the shooting method.
We set $l_4=1$ and give the initial conditions read from (\ref{asym}) at
$\rho^2=5000$ and solve the equation of motion toward $\rho=0$ with
changing $v$ and $m$ to find regular solutions. By using the rescaling
(\ref{cf}), a set of regular solution ($m$,$v$) for $l_4=1$ gives
another one for $l_4$ with ($m/l_4$, $v/l_4^3-m\ln l_4/l_4^3$). When 
$m\lsim 1.4/l_4$, we find the regular solution 
%%%crosses 
ends at the horizon (Fig.\ref{fig7}).
% in Appendix~\ref{B}). 
We show the
numerical results of the sets of $(v,m)$ which give regular solutions
denoted by dots in Fig.\ref{fig1} for $l_4=l_5=1$
case. $v$ is an increasing function of $m$ and it is nontrivial if
finite counter terms can cancel the increasing behaviour of $v$. 
The numerical data is best fit\footnote{
We obtain the fit using the data at $m=40$ and $m=50$.
} with $1.0\times m\ln m$
which is shown in Fig.\ref{fig1} as a line, and then we obtain
$c_1=-1$, $c_2=0$ and $c_4=0$ in our
parameters (\ref{abc}). (The parameter $c_3$ would be fixed similarly
from minimising the on-shell action.)
%%%%%%%%%%%%%%%%%%
%%%  modified  %%%
%%%%%%%%%%%%%%%%%%
Although the line in Fig.\ref{fig1} fits the data very well, the line
does not completely fit them and the small difference between the line
and the dotted data becomes our prediction of chiral condensate which is
shown in Fig.\ref{fig2} with the unit $2T_7\Omega_3=1$.
Since $y(\rho)=y'(\rho)=0$ is a solution, there is no quark
condensate in the chiral limit. The behaviour of chiral condensate is
very similar to that in the large N gauge theory with flavor at finite
temperature studied from $AdS_5$
Schwartzschild black hole~\cite{Babington:2003vm}. Therefore we have
properly picked up finite temperature effects in our D7-brane embedding.
Some comments are in order: 1) When a D7-brane 
%%%crosses 
ends at the horizon, it is nontrivial whether the horizon
is stable under perturbations. However since this is the $AdS$ horizon,
we
expect the horizon is stable under perturbations. In fact we will study
perturbations around the solution in the next section and show there are
no tachyonic modes. 2) We find multiple solutions in the region 
$1.38\leq ml_4\leq 1.41$ (see Fig.\ref{fig8})
% in Appendix~\ref{B}) 
as are
found in~\cite{Kruczenski:2003uq}. This might imply a phase transition
at $m\sim 1.4/l_4$ as studied in QCD at the finite 
temperature~\cite{Kruczenski:2003uq}.
%%%%%%%%%%%%%%%%%%
%%%  modified  %%%
%%%%%%%%%%%%%%%%%%
3) Although the final result in Fig.\ref{fig2} is very similar to that
in~\cite{Babington:2003vm}, the geometries in this paper and
in~\cite{Babington:2003vm} are different. We can see this from studying
the Ricci scalar induced on D7 brane for example. In both cases, the
chiral condensate is zero for the case $m=0$ and the configuration of D7
brane is straight. In this case we can explicitly show that the value of
Ricci scalar is different with each other.

%Before
%introducing the probe D7 brane, the goemetry in this paper is just
%$AdS_5$ and the $N=4$ supersymmetry is kept unbroken which is on the
%contrary broken in the case of $AdS_5$ black hole. Since we study the
%situation in which the probe approximation is a good approximation, the
%backreaction of D7 brane is small and the geometries are still
%different. 

\section{Meson spectrum}

We have not shown there are no tachyonic modes in local excitations
around a regular solution. In the gauge theory side, the fluctuations
about a D7-brane configulation correspond to meson
fields~\cite{Kruczenski:2003be}. In this section we thus aim to study
the meson mass spectrum and see if there are tachyonic modes and if the
meson spectrum captures finite temperature effects.

We first discuss what are expected from the results in the previous
section. Since there is no chiral symmetry breaking in the chiral limit,
it is consistent that we have a mass gap in the spectrum contrary to QCD
at low temperature. The spectrum of heavy masses, much
heavier than $1/l_4$, should be approaching to the mass spectrum in
supersymmetric case plus thermal corrections. With keeping these in
mind, we now solve linearlized equations of
motion. The small (s-mode) fluctuations along $z$ direction\footnote{
It is easily checked the small fluctuations along $y$ direction satisfy
the same equations of motion.} 
are included in the induced metric
\begin{eqnarray}
 ds^2 &=&r^2\left(1-\frac{1}{4r^2l_4^2}\right)^2dx_{dS4}^2 
  +\frac{1}{r^2}(d\rho^2+\rho^2d\Omega_3+y'(\rho)^2 d\rho^2
  +\partial_a z\partial_b zdx^adx^b)
\end{eqnarray}
and the action up to the quadratic terms in terms of $z$ becomes
\begin{eqnarray}
 S &=&-T_7\Omega_3\int d^4xd\rho
  \left(1-\frac{1}{4r_0^2l_4^2}\right)^4\rho^3\sqrt{1+y'(\rho)^2}
  \left\{
   1+\left(1-\frac{1}{4r_0^2l_4^2}\right)^{-2}
   \frac{g_{dS4}^{\mu\nu}\partial_\mu z\partial_\nu z}{2r_0^4}
  \right.
  \nonumber\\
 &&\qquad 
  \left. +\left(1-\frac{1}{4r_0^2l_4^2}\right)^{-1}\frac{z^2}{r_0^4l_4^2}
   +\frac{(\partial_\rho z)^2}{2(1+y'(\rho)^2)}
   \right\}
\end{eqnarray}
where $r_0^2=\rho^2+y(\rho)^2$ and $y(\rho)$ is the regular
solution obtained in the previous section. Then it is easily seen that
the energy carried
by $z$ (s-mode) fluctuations is positive definite and we do not have
tachyonic modes. Notice that the counter terms $S_c$ do not play any
rules in this discussion. Since our main object in this section is
the meson spectrum, we proceed to solve the equations of motion for
$z$. 

The linearized equation of motion for a Kaluza-Klein (KK) mode
$z=\phi_M(x)f(\rho)$ with the four dimensional mass $M$ becomes
\begin{eqnarray}
 \partial_\rho^2 f(\rho) &=&
  \left\{-\frac{3}{\rho}
  +\frac{y'(\rho)y''(\rho)}{1+y'^2(\rho)}
  -\frac{2(\rho+y(\rho)y'(\rho))}{r_0^4l_4^2}
  \left(1-\frac{1}{4r_0^2l_4^2}\right)^{-1}
  \right\}\partial_\rho f(\rho)
  \nonumber\\&&
  +(1+y'(\rho)^2)\left(1-\frac{1}{4r_0^2l_4^2}\right)^{-1}
  \left\{ \frac{2}{r_0^4l_4^2} 
   -\left(1-\frac{1}{4r_0^2l_4^2}\right)^{-1}\frac{M^2}{r_0^4}
  \right\}f(\rho) .
\end{eqnarray}
After normalizing properly, this Kaluza-Klein mode then satisfies 
\begin{eqnarray}
 S&=&-\int d^4x \sqrt{-g_{dS4}}
  (g_{dS4}^{\mu\nu}\partial_\mu \phi_M\partial_\nu \phi_M +M^2\phi_M^2)
  .
\end{eqnarray}

\begin{figure}
 \begin{minipage}{0.5\hsize}
 \begin{minipage}{0.95\hsize}
  \hspace{-2.5ex}
   \includegraphics[width=90mm]{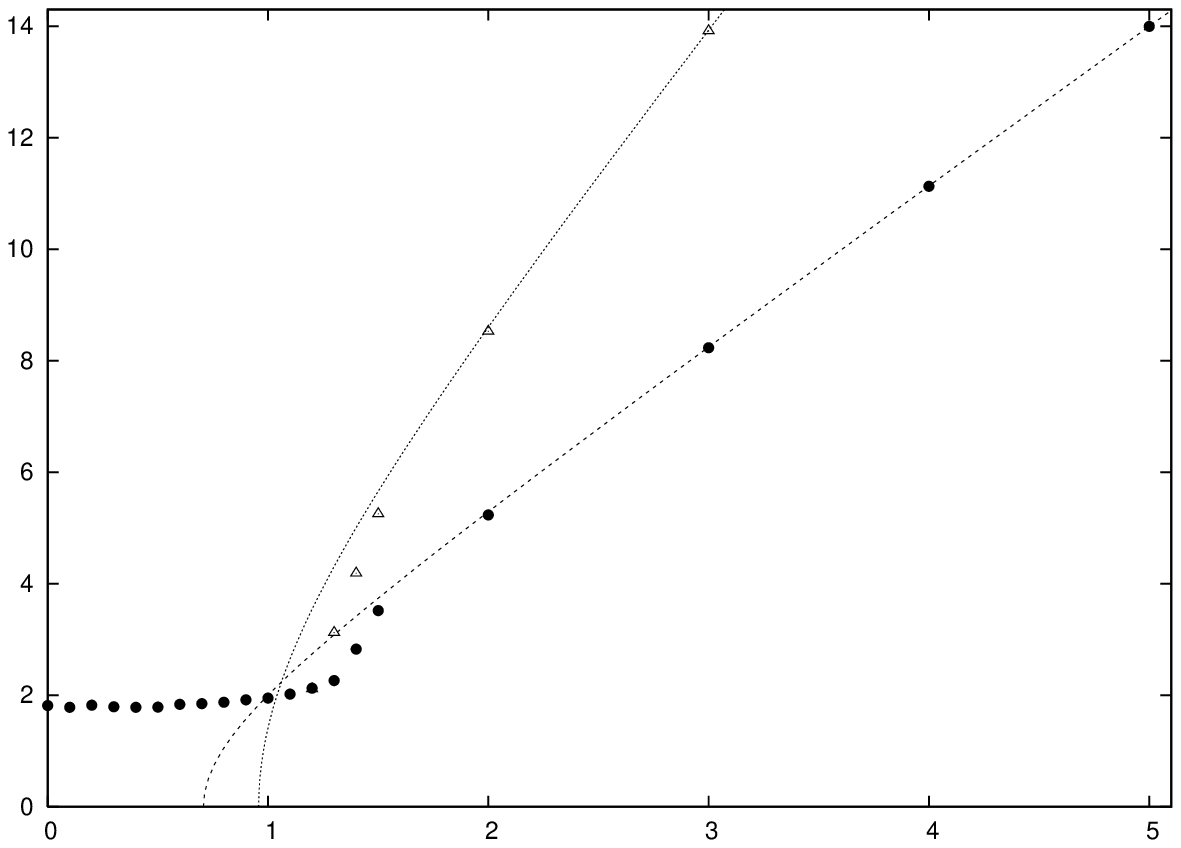}
  \vspace{-4ex}
  \hspace*{2.5ex}
 \caption{The lowest and second KK (meson) masses $M$ vs quark mass
 $m$. The asymptotic forms $M=\sqrt{8m^2-4/l_4^2}$ and
 $M=\sqrt{24m^2-22/l_4^2}$ are given as lines. $l_4=1$.}
  \label{fig3}
 \end{minipage}
 \end{minipage}
 \begin{minipage}{0.5\hsize}
 \begin{minipage}{0.95\hsize}
  \vspace{-3.25ex}
  \hspace{-2.5ex}
   \includegraphics[width=90mm]{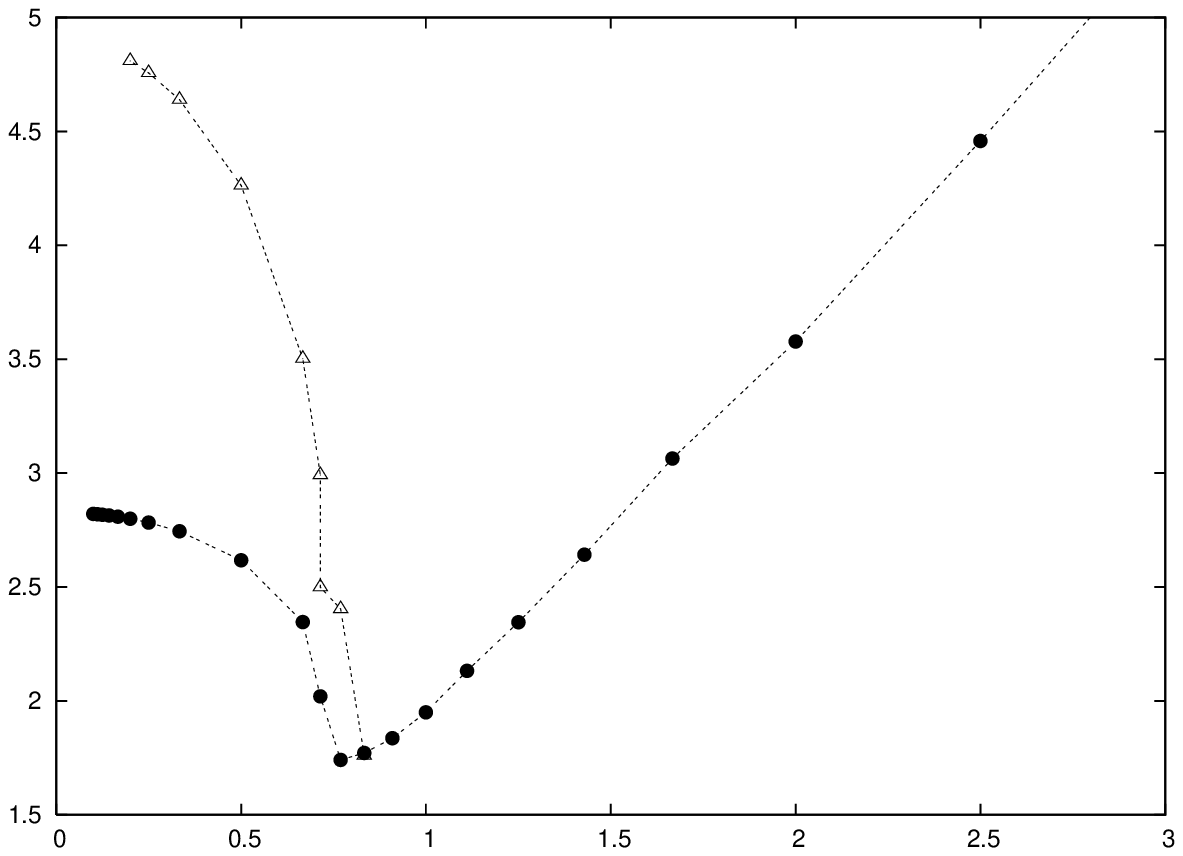}
  \vspace{-4ex}
 \caption{The first and second KK (meson) masses $M$ vs temperature
  $1/l_4$ with fixing quark mass $m=1$. (This is obtained from
  Fig.~\ref{fig3} using the rescaling~(\ref{cf}).)}
  \label{fig9}
 \end{minipage}
 \end{minipage}
\end{figure}

The asymptotic solutions near the $AdS$ boundary are same as those for
$y$ and so $f(\rho)$ should be proportional to $1/\rho^2$ near
the boundary. We give this boundary
condition at $\rho^2=5000$ and solve the equation numerically toward
$\rho=0$. Since $f(\rho_\infty)\propto 1/\rho^2_\infty\rightarrow 0$ in
the $\rho_\infty\rightarrow\infty$ limit, the boundary action $S_c$ does
not change the equation of motion 
%%%or 
and the boundary condition.
%%%%%%%%%%%%%%%%%%
%%%  modified  %%%
%%%%%%%%%%%%%%%%%%
This simply means that the mass of meson field is a physical quantity
and does not depend on a regularization scheme.
The regular solution is only allowed for discrete mass $M$.
For the supersymmetric case, the meson spectrum is calculated
$M^2=4m^2(n+1)(n+2), n=0,1,\cdots$. We have numerically obtained how the
masses of the lowest and the second KK modes ($n=0,1$) change as $m$
changes for the case $dS_4$ radius $l_4=1$. We
show our numerical results by dots in
Fig.\ref{fig3} with $l_4=1$. Using the rescaling, we can obtain the KK
(meson) masses for general $l_4$.
%\begin{figure}
%  \begin{center}
%   \includegraphics[width=120mm]{M.eps}
%  \end{center}
%  \vspace{-3ex}
% \caption{The lowest and second KK (meson) masses $M$ vs quark mass
% $m$. The asymptotic forms $M=\sqrt{8m^2-4/l_4^2}$ and
% $M=\sqrt{24m^2-22/l_4^2}$ are given as lines. $l_4=1$.}
%  \label{fig3}
%\end{figure}
For a large $m\gg 1/l_4$, i.e. a low temperature, we find the asymptotic
form for the mass of lowest KK mode 
\begin{eqnarray}
 M^2=8m^2-4l_4^{-2}
  \label{meson}
\end{eqnarray}
which we put as a line in Fig.\ref{fig3}. Similarly the mass of second
KK mode is given $M^2=24m^2-22l_4^{-2}$.
%%% We do not know the 
The negative 
sign of finite temperature effect in this region is consistent with the
results of QCD at finite temperature~\cite{Roberts:2000aa}.
%%%%%%%%%%%%%%%%%%
%%%  modified  %%%
%%%%%%%%%%%%%%%%%%
This is much clear in Fig.~\ref{fig9} 
%of Appendix~\ref{B} 
where we show
the temperature dependence of meson masses which is obtained from
Fig.~\ref{fig3} using the rescaling~(\ref{cf}) and the decreasing of
meson masses in a low temperature region is due to the negative sign.
This figure is an analog of Fig.4.4 in~\cite{Roberts:2000aa} and as long
as qualitatively they behave similarly. 
%%%  since our dual field
%%% theory is not a real QCD on de-Sitter space.
Although the field theory in our case is different from QCD, this
similarity supports that the dual field theory in our embedding of D7
brane is a field theory on $dS_4$ space.

%We give a figure on the
%temperature dependence of meson masses in Fig.~\ref{fig9} of
%Appendix~\ref{B}, which is an analog of Fig.4.4
%in~\cite{Roberts:2000aa}, and as long as qualitatively they behave
%similarly. 

The mass starts deviating from this asymptotic forms when $m$ becomes
equals to or smaller than $1.4/l_4$
where a D7-brane 
%%%starts crossing 
ends at the horizon, and the KK modes degenerate
below $(1.2\sim 1.4)/l_4$\footnote{In Fig.\ref{fig3} the degeneracy
starts at $1.2/l_4$, but there are numerical errors.
} in our numerical analysis. This again might signal the same phase
transition in QCD at finite temperature studied
in~\cite{Kruczenski:2003uq}. Although in this region we
have numerical errors $0.1\sim 1$ in the lowest KK mass square, the mass
square never crosses zero. This is supported from the
fact that the $AdS$ horizon is stable. 
%%%%%%%%%%%%%%%%%%
%%%  modified  %%%
%%%%%%%%%%%%%%%%%%
The flatness of the mass is understood as the large part of mass comes
from thermodynamical fluctuations when the temperature 
$T=2\pi/l_4$ becomes larger than the bare quark mass $m$. This
behaviour is clear in Fig.~\ref{fig9} 
%of Appendix~\ref{B} 
in which the mass increases
linearly as the temperature increases in a high temperature region.
These all behaviours are very
similar to those in a large N gauge theory with flavor at finite 
temperature~\cite{Babington:2003vm, Kruczenski:2003uq}. 

In this section we did not study fluctuations of gauge fields on a
D7-brane which correspond to vector mesons in the gauge theory side. It
is known that
there is a mass gap in supersymmetric case~\cite{Kruczenski:2003be}
and thus we expect a mass gap in our system at least when 
$m\gg 1/l_4$. It is nontrivial whether we have tachyonic modes in vector
mesons when $m$ becomes smaller. We leave this question as a future
work.

\section{Wilson loops}

In previous sections, we have studied the chiral condensate and meson
spectrum and have obtained consistent results with those of a large N
gauge theory with flavor at a finite
temperature. In this section we study Wilson loops and the static quark
anti-quark potential. In the $AdS$ dual picture, the Wilson loop is
described by a string lying along a geodesic in the $AdS$ bulk with the
endpoints on the boundary or a 
D7-brane~\cite{Rey:1998ik, Kruczenski:2003be}, which in our
situation is the $dS_4$ hypersurface. The end points
represent the positions of quark and anti-quark on a D7-brane. The
Nambu-Goto action becomes
\begin{eqnarray}
 S=\frac{T\sqrt{-g_{tt}}}{2\pi}\int dx\sqrt{
  (u^2-l_4^{-2})^2g_{xx}+(\partial_x u)^2}
\end{eqnarray}
where we have used the coordinate system in (\ref{met2}) and $u$ is a
function of $x$, i.e. $u=u(x)$. $g_{tt}$ and
$g_{xx}$ are $(t,t)$ and $(x,x)$ components of 4d de-Sitter
metric. We assume the time scale $T$ is much smaller than $l_4$ and
then we approximate $g_{tt}$ and $g_{xx}$ are constant ($g_{xx}=1$).
In this paper we only consider the case $m\gg 1/l_4$ and further
approximate the D7-brane configuration is straight $y(\rho)=m$. The
open string which stretches in the $AdS$ bulk with endpoints at $u=m$
conserves the following quantity $c$,
\begin{eqnarray}
 \frac{(u^2-l_4^{-2})^2}{\sqrt{
  (u^2-l_4^{-2})^2+(\partial_x u)^2}} = c =u_c^2-l_4^{-2}
\end{eqnarray}
and $\sqrt{c}$ measures the distance between the horizon $u=1/l_4$ and
the point $u_c$ until where a string extends from $u=m$. Thus when $c$
goes to zero, the quark and anti-quark are no longer connected by a
string and move freely.
The distance $L$ between two heavy quarks is obtained from the above
equation,
\begin{eqnarray}
 L=\frac{2}{\sqrt{c}}\int^{m/\sqrt{c}}_{u_c/\sqrt{c}}dy
  \frac{1}{(y^2-c^{-1}l_4^{-2})\sqrt{(y^2-c^{-1}l_4^{-2})^2-1}} .
\end{eqnarray}
The energy of Wilson loop is calculated from the on-shell Nambu-Goto
action,
\begin{eqnarray}
 E &=& 2\sqrt{c}
   \int_{u_c/\sqrt{c}}^{m/\sqrt{c}} dy 
   \frac{(y^2-c^{-1}l_4^{-2})}{\sqrt{(y^2-c^{-1}l_4^{-2})^2-1}} .
\end{eqnarray}
Here we have dropped various numerical factors which are not important
in our discussions.
It is easily seen that $L$ ($E$) is given by a function of $cl_4^2$
times $c^{-1/2}$ ($c^{1/2}$). Therefore as before we can set $l_4=1$
without losing generality. 

We divide $L$ into three regions; (1) the short distance
region $L\ll 1/m$, (2) the intermediate distance region $1/m\ll L \lsim l_4$
and (3) the large distance region $L\gsim l_4$.

\begin{figure}
 \begin{minipage}{0.5\hsize}
 \begin{minipage}{0.95\hsize}
  \vspace{-3ex}
  \hspace{-3ex}
   \includegraphics[width=90mm]{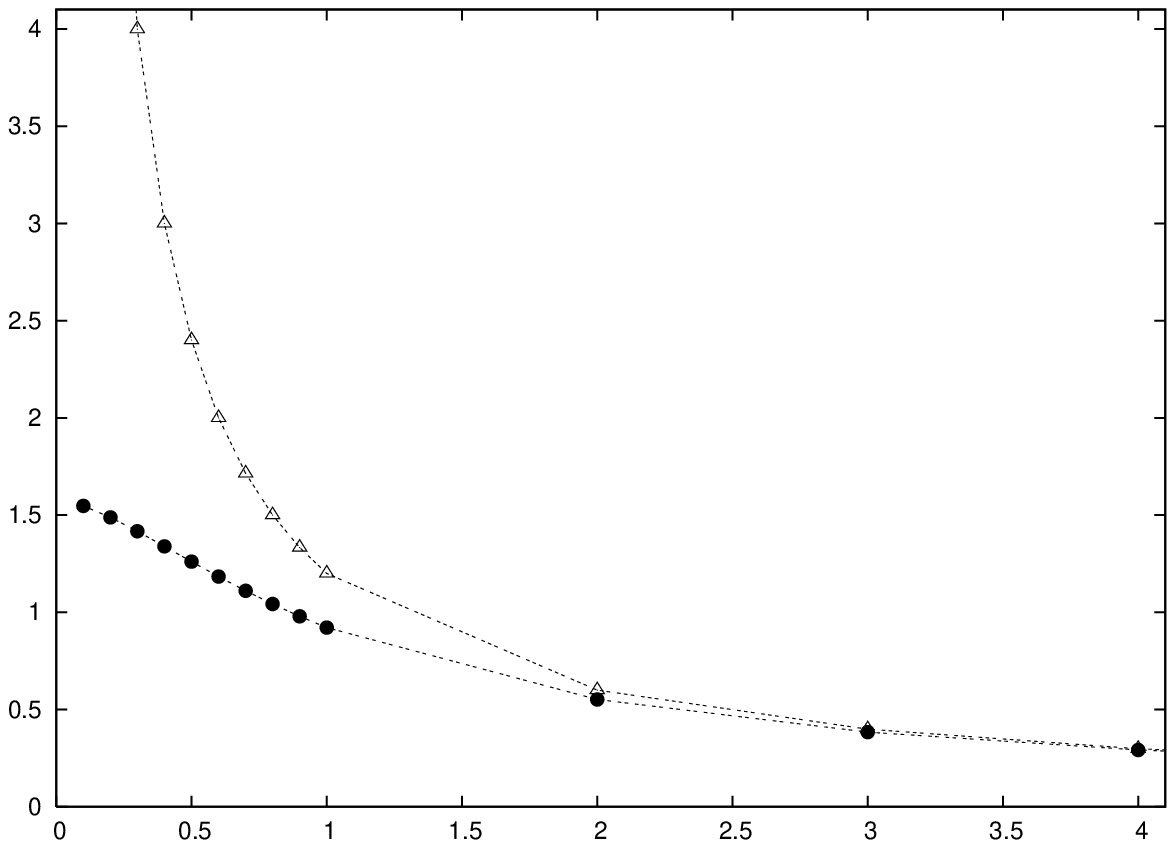}
  \vspace{-3ex}
 \caption{The distance $L$ vs $\sqrt{c}$. The
  upper (bottom) curve is for $l_4\rightarrow\infty$ ($l_4=1$).
  $m=20$.}
  \label{fig5}
 \end{minipage}
 \end{minipage}
 \begin{minipage}{0.5\hsize}
 \begin{minipage}{0.95\hsize}
  \hspace{-2.5ex}
   \includegraphics[width=90mm]{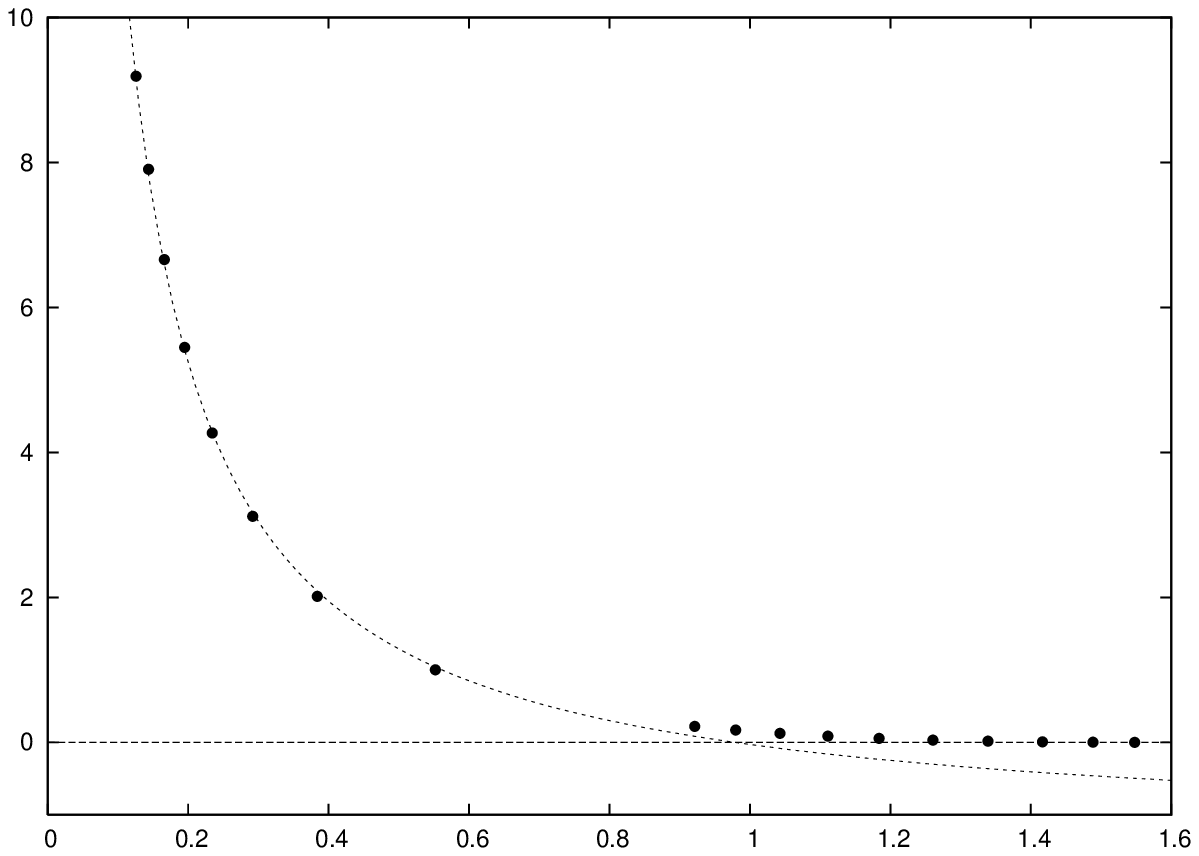}
  \vspace{-3ex}
 \caption{The binding energy $E-2m_q$ vs $L$.
  %The line is $1.42/L-1.76/l_4$.
  The line is $1.32/L-1.35/l_4$.
  $E$ exactly goes to zero at $L=L_{max}$. $m=20$. $l_4=1$.}
  \label{fig6}
 \end{minipage}
 \end{minipage}
\end{figure}

We compute $L$ and $E$ numerically for (2) and (3) regions and show
those results by dots in Fig.\ref{fig5} and Fig.\ref{fig6} where we put
$l_4=1$ and took $m=20$.
In Fig.\ref{fig5}, there are two lines with dots. The upper curve is the
distance $L$ for the supersymmetric case ($l_4\rightarrow\infty$) and
the lower curve is for $l_4=1$. In the region (2) ($L\lsim l_4$), they
approach with each other, and the binding energy of quark and anti-quark
bound state $E-2m_q$ is well fit\footnote{
We use the data at $0.1\leq L\leq 0.4$.}
with a Coulomb type potential
\begin{eqnarray}
 E-2m_q &=& \frac{1.3}{L}-\frac{1.3}{l_4}, \hspace{3ex}
  (m_q=m-l_4^{-1})
\end{eqnarray}
which is given as a line in Fig.\ref{fig6}. The quark mass is identified
as $m-l_4^{-1}$ since a string can extend up to $u=l_4^{-1}$ not
$u=0$~\cite{Rey:1998bq}. The difference between $m_q$ and $m$
obtained from the AdS/CFT dictionary~\cite{Gubser:1998bc} could be
understood as a renormalization group effect since the endpoints are
located at $\rho=0$ on a D7-brane, not at $\rho=\infty$ and we should
use $m$ at a short distance region, i.e. $L\ll 1/m$.
The $dS_4$
curvature correction $-1.3/l_4$ reduces the binding energy which is
consistently understood as the finite temperature effect.

We find discrepancies from N=2 SYM when $c$ goes to zero (region
(3)). The distance
$L$ has the maximal length $L_{max}$ for $dS_4$ case
\begin{eqnarray}
 L_{max} &\sim&1.6\times l_4 .
\end{eqnarray}
The binding energy goes to zero at $L_{max}$ (faster than a Coulomb type
potential) from the above (Fig.\ref{fig6}) and
they become free particles when $L$ is larger than $L_{max}$.
%\begin{eqnarray}
% E &=& 1.15\times (\frac{1}{L}-0.64)^2, \hspace{3ex}
%  \left(\frac{1}{L}\lsim \frac{1.2}{l_4}\right)
%\end{eqnarray}
A similar result was obtained from $AdS$ black holes at finite
temperature~\cite{Rey:1998bq}.

We compute $E$ and $L$ in the $L\rightarrow 0$ limit
(region (1)) and obtain 
\begin{eqnarray}
 E &=& (m^2-l_4^{-2})L .
\end{eqnarray}
This is the linear potential and the finite temperature effect appears
as reducing the string tension as expected. (Here we chose the bare mass
$m$ as a quark mass, since $L\rightarrow 0$ is the high energy limit.)

\section{Conformal transformation}

With the success of producing finite temperature effects from our
D7-brane embedding, we would like to try answering why we have obtained
such reasonable results. As briefly discussed in Introduction, the
different slice of $AdS$ space give different results on the CFT side. 

We study a bulk scalar field with mass $M$.
\begin{eqnarray}
 S &=&\int d^5x\sqrt{-g_{AdS}}\left(-(\partial\phi)^2-M^2\phi^2\right) .
\end{eqnarray} 
Since the components of two metrics (\ref{met1}) and (\ref{met2}) are
same at the leading order near the
$AdS$ boundary, the asymptotic solutions of static configuration along
$dS_4$ is given
\begin{eqnarray}
 \phi(u)&=&c_1u^{\alpha_+}+c_2u^{\alpha_-}, \hspace{3ex}
  \alpha_\pm = -2\pm\sqrt{4+M^2l_5^2},
\end{eqnarray}
where $c_1$ and $c_2$ are integration constants. Here we only discuss
the case $-4\leq M^2l_5^2< -3$ otherwise the first correction to
$c_1u^{\alpha_+}$ asymptotic solution is larger than $c_1u^{\alpha_-}$.
(But the essential arguments below are still same for general $M$.)
In terms of 4d Minkowski coordinate system, $u=Rt/l_4$
(Appendix~\ref{A}),
\begin{eqnarray}
 \phi(u)&=&c_1(Rt/l_4)^{\alpha_+}+c_2(Rt/l_4)^{\alpha_-} .
\end{eqnarray}
Using the original AdS/CFT correspondence, the scaling dimension of
corresponding CFT operator ${\cal O}(x)$ is read from this
solution and is $4+\alpha_+$.
We introduce the cutoff at a constant $R=R_\infty$ and use the AdS/CFT
dictionary. From straightforward calculations, we
obtain the time dependent source and vacuum expectation value,
\begin{eqnarray}
 {\cal L} &=& {\cal L}_{M4} + c_1(t/l_4)^{\alpha_+}{\cal O}(x),
  \\
 \langle {\cal O}(x) \rangle &\propto&
  \lim_{R_\infty\rightarrow\infty}
  \frac{\delta S}{(Rt/l_4)^{\alpha_+}\delta c_1}\Big|_{c_1\rightarrow 0}
  \propto c_2(t/l_4)^{\alpha_-},
\end{eqnarray}
where ${\cal L}_{M4}$ is the CFT action on 4d Minkowski space,
$x=(t,x_i)$. (Notice that the on-shell action from
$t\rightarrow\pm\infty$ goes to zero compared to that from the cut
$R=R_\infty$ as the cut approaches to the $AdS$ boundary.) 
From the D3-brane effective theory's point of view on Minkowski space,
we are considering responses against time dependent sources. As
discussed in Section~2, the scalar field on D7-brane corresponds to the
quark anti-quark composite operator. Therefore we were studying a time
dependent mass term for quarks (thus supersymmetry is broken) and
responses to that.

On the other hand, the (A)dS/dS correspondence claims that introducing
the cutoff at constant $u=u_{\infty}$ and applying the AdS/CFT
dictionary. We then obtain
\begin{eqnarray}
 {\cal L} &=& {\cal L}_{dS4} + c_1{\cal O}(x),
  \\
 \langle {\cal O}(x) \rangle &\propto&
  c_2,
\end{eqnarray}
where ${\cal L}_{dS4}$ is the CFT action on $dS_4$, $x=(s,x_i)$
(Appendix~\ref{A}), and the 
vacuum expectation value is calculated around $dS_4$ background.
The operator ${\cal O}$ is same since the asymptotic solution is same.

The cut near the $AdS$ boundary corresponds to the UV cutoff in the CFT
side. The above result indicates a different UV regularization induces
two different results on correlation functions and the difference does
not seem due to just a different scheme. Let us try to give some
explanation on this point. We again study the on-shell value of
action. If we keep $c_1$ nonzero and compute the on-shell action, we
obtain in each case
\begin{eqnarray}
 S &=& \int d^4x\sqrt{-g_{M4}} \frac{R^5}{l_5^5}\phi\partial_R\phi
  \Big|^{R=R_\infty}
  =\int d^4x\sqrt{-g_{M4}}
  \frac{\alpha_+ c_1^2t^{2\alpha_+}}{l_5^5l_4^{2\alpha_+}}
  R^{4+2\alpha_+}_{\infty} +\cdots ,
  \label{S1}\\
 S &=& \int d^4x\sqrt{-g_{dS4}} 
  \left(\frac{u^2}{l_5^2}-\frac{l_5^2}{l_4^2}\right)^{5/2}
  \phi\partial_u\phi
  \Big|^{u=u_\infty}
  =\int d^4x\sqrt{-g_{dS4}}
  \frac{\alpha_+ c_1^2}{l_5^5}
  u^{4+2\alpha_+}_{\infty} +\cdots .
\end{eqnarray}
(Here we ignored the possible contributions from $t=\pm\infty$.)
The leading contribution is divergent and takes a different value
except when $t\sim s\sim l_4$. Thus they are not related just by a
finite renormalization. 

Because of a time dependent coupling, we have time dependent divergences 
(\ref{S1}). Thus 
we may mix time $t$ and the UV cutoff $R_{\infty}$ and define a new
time $s$, and similarly define a new UV cutoff $u_{\infty}$ such that
the time dependence disappears. (This is nothing but a coordinate
transformation from $AdS$ point of view (\ref{ts}).) Taking the limit
$R_{\infty}\rightarrow\infty$, the mixture becomes zero and the new
time $s$ is still $t$ (and so the metric unchanged). Since $dS_4$ is
conformal to flat, we act the conformal transformation (and scale
transformation to fields) and change the
metric from Minkowski to $dS_4$,
\begin{eqnarray}
 g_{M4}=\Omega^{-2}g_{dS4}, \hspace{3ex}
  \Omega =\frac{l_4}{s}=\frac{l_4}{t} .
\end{eqnarray}
Since N=4 SYM, ${\cal L}_{M4}(\Phi)$ ($\Phi$ represents the fields in
the CFT), is scale invariant, the resultant action ${\cal L}_{dS4}$
becomes ${\cal L}_{dS4}={\cal L}_{M4}(\Phi)
+{\cal L}(\partial\Omega,\Omega,\Phi)$ and the contraction of Lorentz
indices is given by $dS_4$ metric. (We do not consider anomaly here.)
The second part is the terms which always include the (time) derivative
of $\Omega$\footnote{
The D3-brane action might have couplings to gravity,
when gravity is coupled, which are consistent with superconformal 
invariance, such as $R\phi^2/6$ for a scalar field $\phi$ and $R$ is the
scalar curvature of the background 4d metric~\cite{Seiberg:1999xz}. Then
the second part is written with a background gravity in a general
covariant way.
}. Since the scaling dimension of ${\cal O}$ is $4+\alpha_+$,
$c_1(t/l_4)^{\alpha_+}{\cal O}(x)$ part in Lagrangian is
transformed into $c_1{\cal O}(x)$ and the vacuum expectation value
$\langle {\cal O}(x) \rangle \propto c_2(t/l_4)^{\alpha_-}$ is also
transformed into $\langle {\cal O}(x) \rangle \propto c_2$. Although
this argument is quite unsatisfactory, the coupling constant
and vacuum expectation value in CFT on Minkowski is conformally
transformed into those in CFT on de-Sitter space. The cutoff on
Minkowski space is also mapped into the cutoff on de-Sitter space.
We hope this is touching something behind the (A)dS/dS correspondence
and why we obtained reasonable results.

\section{Conclusions and Discussions}

We have studied a D7-brane probe in $AdS_5\times S^5$ geometry.
We identify the normal direction to a $dS_4$ hypersurface in
$AdS_5$ as the energy scale in the dual gauge theory and obtain a
D7-brane configuration which is static along the $dS_4$ directions. The
scalar field on a D7-brane is a
bulk field of $AdS_5$ foliated by $dS_4$ and can be treated as a bulk
field in the Randall Sundrum model with non fine-tuned
brane(s) or in the (A)dS/dS correspondence studied by Alishahiha
et al.~\cite{Alishahiha:2004md, Alishahiha:2005dj}.
Thus the field theory dual is expected to be a CFT with flavor on
$dS_4$. We have checked this expectation is true by identifying the
finite temperature effects in the chiral condensate, meson mass spectrum
and quark anti-quark potential and showing they are properly reproduced.
Therefore we realize the gravity dual of non supersymmetric field theory
by breaking supersymmetry by a probe brane.

The regularized on-shell action is log divergent in
terms of m, which is the asymptotic locus of the D7-brane, and thus
there is an infinite potential barrier between the $dS$ embedding and
supersymmetric embedding. This might indicate that different UV
regularizations induce renormalization flows into different IR physics
which are not connected by perturbative modes or
finite renormalizations, but might be transformed with each other by a
complicated (conformal) transformation.

We showed there is no tachyonic mode in the perturbations of scalar
fields localized on the D7-brane. We expect there is no tachyonic mode
in the perturbations of
gauge fields on the D7-brane as well, since there is a mass gap in
the supersymmetric limit. When we include the back reaction of
D7-brane, the gravitational solution is expected to be still
similar. Thus this realization of de-Sitter space on the D7-brane might
be phenomenologically
interesting, since the possibility of the localized
gravity in D3/D7 system is discussed in~\cite{LiamFitzpatrick:2005wq}. 

We have little knowledge on the quantum field theory on de-Sitter
space. Because of numerical error, we did not pay much attention to the
high temperature regions. It might be interesting to study them and
phase transitions by improving the accuracy of numerical analysis. It
might be also interesting to study two point or multi point functions in
this direction.

\section*{Acknowledgement}

%Thank you.

The author would like to thank Shigeki Sugimoto for reading the
manuscript and giving useful comments. The author would like to
appreciate the free atmosphere at Shanghai Institute for Advanced
Studies.

\appendix

\section{Minkowski and de-Sitter slices in Anti de-Sitter space}
\label{A}

The $AdS_5$ with the curvature radius $l_5$ is a hypersurface
\begin{eqnarray}
 X_0^2+X_5^2-X_i^2-X_4^2=l_5^2, \hspace{3ex} i=1,2,3,
\end{eqnarray}
in the 6 dimensional flat space with the metric
\begin{eqnarray}
 ds^2 &=& -dX_5^2+dX_i^2+d(X_4+X_0)d(X_4-X_0) .
\end{eqnarray}
The 4d Minkowski space with coordinates $(t,x_i)$ is embedded as
follows,
\begin{eqnarray}
 X_5=\frac{Rt}{l_5},\hspace{3ex}
 X_i=\frac{Rx_i}{l_5},\hspace{3ex}
 X_4-X_0=-R,\hspace{3ex}
 X_4+X_0=\frac{l_5^2}{R}+\frac{R}{l_5^2}(\vec{x}^2-t^2) ,
%  \\
% X_0&=&\frac{l_5^2}{2R}
%  \left(1+\frac{R^2}{l_5^4}(l_5^2+\vec{x}^2-t^2))\right),\hspace{3ex}
% X_4=\frac{l_5^2}{2R}
%  \left(1-\frac{R^2}{l_5^4}(l_5^2-\vec{x}^2+t^2))\right)
\end{eqnarray}
and the $AdS_5$ metric is given
\begin{eqnarray}
 ds^2&=&\frac{R^2}{l_5^2}(-dt^2+d\vec{x}^2)+\frac{l_5^2}{R^2}dR^2 .
\end{eqnarray}
Similarly the $dS_4$ space with coordinates $(s,x_i)$ is embedded along
$X_0, X_i$ and $X_4$ satisfying
\begin{eqnarray}
 X_0^2-X_i^2-X_4^2=-l_4^2
  \left(\frac{u^2}{l_5^2}-\frac{l_5^2}{l_4^2}\right) .
\end{eqnarray}
The embedding is
\begin{eqnarray}
 X_5&=&\frac{l_4}{l_5}u, \hspace{3ex}
 X_i=\frac{l_4x_i}{s}\sqrt{\frac{u^2}{l_5^2}-\frac{l_5^2}{l_4^2}},
 \hspace{3ex}
 X_4-X_0= \frac{l_4}{s}\sqrt{\frac{u^2}{l_5^2}-\frac{l_5^2}{l_4^2}},
 \\
 X_4+X_0&=&l_4(s-\frac{\vec{x}^2}{s})
 \sqrt{\frac{u^2}{l_5^2}-\frac{l_5^2}{l_4^2}},
% \\
% X_0&=&\frac{l_4}{2}\left(s-s^{-1}-\vec{x}^2s^{-1}\right)
%  \sqrt{\frac{u^2}{l_5^2}-\frac{l_5^2}{l_4^2}}\\
% X_4&=&\frac{l_4}{2}\left(s+s^{-1}-\vec{x}^2s^{-1}\right)
%  \sqrt{\frac{u^2}{l_5^2}-\frac{l_5^2}{l_4^2}}
\end{eqnarray}
and $AdS_5$ metric becomes
\begin{eqnarray}
 ds^2=\left(\frac{u^2}{l_5^2}-\frac{l_5^2}{l_4^2}\right)
  \left(\frac{l_4^2}{s^2}(-ds^2+d\vec{x}^2)\right)
  +\left(\frac{u^2}{l_5^2}-\frac{l_5^2}{l_4^2}\right)^{-1}du^2 .
\end{eqnarray}
The transformation between two coordinate systems is easily
obtained from the embeddings:
\begin{eqnarray}
 x_i&=&x_i\\
 t=\frac{su}{l_5}
  \left(\frac{u^2}{l_5^2}-\frac{l_5^2}{l_4^2}\right)^{-1/2}
   &\leftrightarrow&
   s=\left(t^2-\frac{l_5^4}{R^2}\right)^{1/2}
   \label{ts}\\
 R=\frac{l_4l_5}{s}
  \left(\frac{u^2}{l_5^2}-\frac{l_5^2}{l_4^2}\right)^{1/2}
  &\leftrightarrow&
 u=\frac{Rt}{l_4}
\end{eqnarray}
The $AdS$ boundary corresponds to $R\rightarrow\infty$ and equivalently
to $u\rightarrow\infty$. Near the boundary, we have $t=s$.

%\pagebreak
%\section{Figures}\label{B}
%\begin{figure}[H]
% \begin{minipage}{0.5\hsize}
% \begin{minipage}{0.95\hsize}
%  \hspace{-3ex}
%   \includegraphics[width=90mm]{curves.eps}
%  \vspace{-3ex}
% \caption{The regular solutions $y(\rho)$ vs $\rho$.
%  From the top $m=1.5,1.43,1.37$ and $1.3$. The horizon radius is
% $0.5$. $l_4=1$.}
%  \label{fig7}
% \end{minipage}
% \end{minipage}
% \begin{minipage}{0.5\hsize}
% \begin{minipage}{0.95\hsize}
%  \vspace{-2.5ex}
%  \hspace{-2.5ex}
%   \includegraphics[width=90mm]{curve2.eps}
%  \vspace{-3ex}
% \caption{The multiple solutions $y(\rho)$ at $m=1.4$ vs $\rho$.
%   $l_4=1$.}
%  \label{fig8}
% \end{minipage}
% \end{minipage}
%\end{figure}
%\begin{figure}[H]
% \begin{minipage}{0.5\hsize}
% \begin{minipage}{0.95\hsize}
%  \hspace{-3ex}
%   \includegraphics[width=90mm]{KK3.eps}
%  \vspace{-3ex}
% \caption{The first and second KK (meson) masses $M$ vs temperature
%  $1/l_4$ with fixing quark mass $m=1$. (This is obtained from
%  Fig.~\ref{fig3} using the rescaling~(\ref{cf}).)}
%  \label{fig9}
% \end{minipage}
% \end{minipage}
%\end{figure}

\end{document}